\documentclass[]{foils}
\usepackage{epsf}
\usepackage[dvips]{graphics}

\date{}
\MyLogo{Marek W. Gutowski, M\c{a}dralin, August 29th -- September 1st,  2002}

\input{emlines.sty}
\begin{document}

\title{INTERVAL METHODS FOR FMR SPECTRA SIMULATION}

\author{Marek W. Gutowski\\
\bigskip
	gutow@ifpan.edu.pl
}

\maketitle

\vfil
{\small
\begin{center}
Part of author's statutory activity in:\\
Institute of Physics\\
Polish Academy of Sciences\\
Warsaw
\end{center}
}

\foilhead{What is interval calculus?}

A branch of (numerical) mathematics, which makes possible to evaluate
{\bf ranges} (bounds) of algebraic expressions over finite domains, not
just for specific values of their parameters.

The results obtained on different computers may differ, i.e. the bounds
are usually overestimated, but {\bf always} include the true result.

\bigskip
Examples:
$$
\left(\left[-2,3\right]\right)^{2}=\left[0,9\right]
$$

$$
\left[-2,3\right] + \left[0,1\right] = \left[-2,4\right]
$$

but, surprisingly

$$
\left[0,1\right] - \left[0,1\right] = \left[-1,+1\right]
$$

\foilhead{What is FMR?}

{\bf \underline{F}erro\underline{m}agnetic \underline{r}esonance} is
the resonance absorption of electromagnetic radiation by magnetic
bodies placed in an external magnetic field.

The magnetization vector of the ferromagnetic body, which is located in
an external magnetic field, is freely precessing around the field
vector with some frequency.  The precession is damped and spontaneously
decays. The relation between the field strenght and the precessing
frequency is simple:
\begin{center}

\fbox{

$\omega=\gamma{\mathbf H}_{eff}$

}
\end{center}

where ${\mathbf H}_{eff}$ is an {\bf effective field\/} acting on the
body, and $\gamma$ is a constant known as {\em gyromagnetic ratio\/}.

${\mathbf H}_{eff}$ includes: {\bf external} magnetic field, {\bf
demagnetizing} field, due to the sample's geometry (shape), and {\bf
anisotropy} field.

\newpage
\foilhead{Two types of measurements}
There are two possible kinds of experiments:
\begin{enumerate}
\item with fixed external magnetic field and variable frequency, and
\item with fixed (microwave) frequency and variable external field
\end{enumerate}

We will talk about the spectra obtained with the second method.  Here
the resonance absorption is observed only for some magnitudes of an
external field, called {\em resonance fields}.  Due to the presence of
various internal fields, the observed resonance fields will be
generally different for different orientations of the sample in the
external field.

\foilhead{The resonance condition}

The resonance condition is usually written in the form of the well
known relation:
\begin{equation}\label{r_freq}
	\left(\frac{\omega}{\gamma}\right)^{2} = \frac{1}
	{\left(M\sin\theta\right)^{2}}
	\left[\frac{\partial^{2}E}{\partial\theta^{2}}
	\cdot\frac{\partial^{2}E}{\partial\varphi^{2}}
	- \left(\frac{\partial^{2}E}{\partial\theta\ \partial\varphi}
	\right)^{2}\right]
\end{equation}
where the second derivatives of the free energy $E({\mathbf H}_{ext},
{\mathbf M}, \ldots)$ are taken {\bf at equilibrium position} of the
magnetization vector ${\mathbf M}$, and $(\theta, \varphi)$ describe
the orientation of this vector in the polar reference frame.

In order to find the equilibrium position of ${\mathbf M}$, one needs
first to solve the system of equations

$$
	\left\{
	\begin{array}{lcr}
	\partial E/\partial\theta &=& 0\\
	\partial E/\partial\varphi &=& 0
	\end{array}
	\right.
$$

and making sure that its solution(s) are indeed at the minimum of the
free energy, i.e. that the r.h.s. of (\ref{r_freq}) is strictly
positive.

\foilhead{Outline of classical calculation method}
\begin{enumerate}
\item fix the orientation and magnitude of the external field
\item find numerically the equilibrium position of magnetization vector
${\mathbf M}$ and verify that it is stable
\item calculate the resonance frequency $\omega$
\item if $\omega$ coincides with the frequency used in experiment then
we have found the resonance field, otherwise the calculations should be
repeated for other value of $H_{ext}$ (with the same orienatation).
\end{enumerate}

The equilibrium position of ${\mathbf M}$ may be hard to find (and thus
inexact), since even in amorphous, i.e. non-crystalline, samples,
usually ${\mathbf M}$ and ${\mathbf H}_{ext}$ {\bf are NOT parallel}.

\newpage
\foilhead{Interval method}

The list of 3D boxes
$\Delta\varphi\times\Delta\theta\times\Delta H_{ext}$ is systematically
reviewed, starting from the single initial box
\begin{center}
$\left[0,2\pi\right]\times\left[0,\pi\right]\times\left[0,H_{max}\right]$
\end{center}
For each box in succession the series of tests are applied, leading
either to elimination of the box from the list or to its splitting into
two (smaller) offspring boxes.  Failing any of the tests below
(answer: {\bf NO}) eliminates the box from list
\begin{itemize}
\item does the interval $\partial E/\partial\varphi$ contains zero?
\item does the interval $\partial E/\partial\theta$ contains zero?
\item does the interval describing r.h.s of relation (\ref{r_freq})
contains positive numbers?
\item does the interval $\omega$, computed from relation (\ref{r_freq})
contains the experimental frequency $\omega_{exp}$?
\end{itemize}

\foilhead{Interval method --- some details}
Boxes not failing applied tests remain on the list.  The largest of
them is then selected and divided into two parts, each of which are
tried again.  We continue this procedure until the list is empty or
contains only small boxes, i.e. in our case
$\Delta\varphi=\Delta\theta\ \le\ 2\cdot 10^{-6}\ {\rm rad} \approx
10^{-4}$~degree, $\Delta H=0.005$~Oe.  The maximum length of list is
usually close to $200$, but occasionally, for some "difficult"
orientations, it exceeds $2000$.  On exit, the small neighboring boxes
are 'glued' together, if necessary.  In rare cases, for some directions
of ${\mathbf H}$, this procedure leads to higher inaccuracy in
determining $H_{res}$.

For amorphous wire, depending on orientation of ${\mathbf H}_{ext}$,
zero, one, two or even more boxes (resonance fields) are returned, see
figures.

Typical running time, for $\theta_{ext} = 0^{\circ}, 2^{\circ},
4^{\circ}, \ldots 180^{\circ}$ is around $12$~min. on a $100$~MHz PC.

\foilhead{Advantages of interval method}
\begin{itemize}
\item no resonance field is ever missed

\item complete elimination of numerical inaccuracies

\item the equilibrium positions are calculated exactly, {\bf without
any simplifications}, even when the anisotropies are quite complicated

\item {\bf reliable} replacement for other methods

\item the method may be easily extended to reliably reconstruct the
values of unknown anisotropy constants and other material parameters,
{\bf together with their uncertainties}, from experimental data, thus
replacing the usual trial-and-error procedures. {\bf In this case no
classical counterpart --- other than guessing --- exists.}

\end{itemize}

\foilhead{Disadvantages of interval method}

\begin{itemize}

\item neglible or non-existent knowledge of interval methods among the
practitioners in the field and, generally, among physicists at
large\footnote{To find more on interval methods click on the URL:\\
\normalsize	{\tt http://www.cs.utep.edu/interval-comp/} }
\item increased requirements for the raw computing power (moderate)
\end{itemize}

\foilhead{List of figures}

All figures simulated with: $\omega_{exp} = 2\pi\times 9.243$~GHz,
$g=2.00$ and $4\pi M_{s}=6400$~Gs.  Anisotropy constants $K_{u}$ and
$K_{4}$ are given in $10^{5}\times$erg/cm$^{3}$, angles ($x$-axis) in
degrees, and resonance fields ($y$-axis) -- in Gs.

\small
\begin{enumerate}
\item No anisotropy at all, $K_{u}=K_{4}=0$
\item $K_{u}=-1.11$, $K_{4}=0$
\item $K_{u}=+0.20$, $K_{4}=-K_{u}$
\item $K_{u}=+2.00$, $K_{4}=+4.41$
\item $K_{u}=+1.59$, $K_{4}=+2.85$
\item $K_{u}=+1.19$, $K_{4}=-2.38$
\item $K_{u}=-0.19$, $K_{4}=+2.90$
\item $K_{u}=-1.19$, $K_{4}=+7.16$
\item $K_{u}=-1.39$, $K_{4}=+3.18$
\item $K_{u}=-5.96$, $K_{4}=-1.19$
\item $K_{u}=-7.16$, $K_{4}=+4.77$
\item $K_{u}=-1.39$, $K_{4}=+2.38$
\end{enumerate}

\newpage

\begin{figure}[h]
\epsfysize=18cm
\centerline{\epsfbox{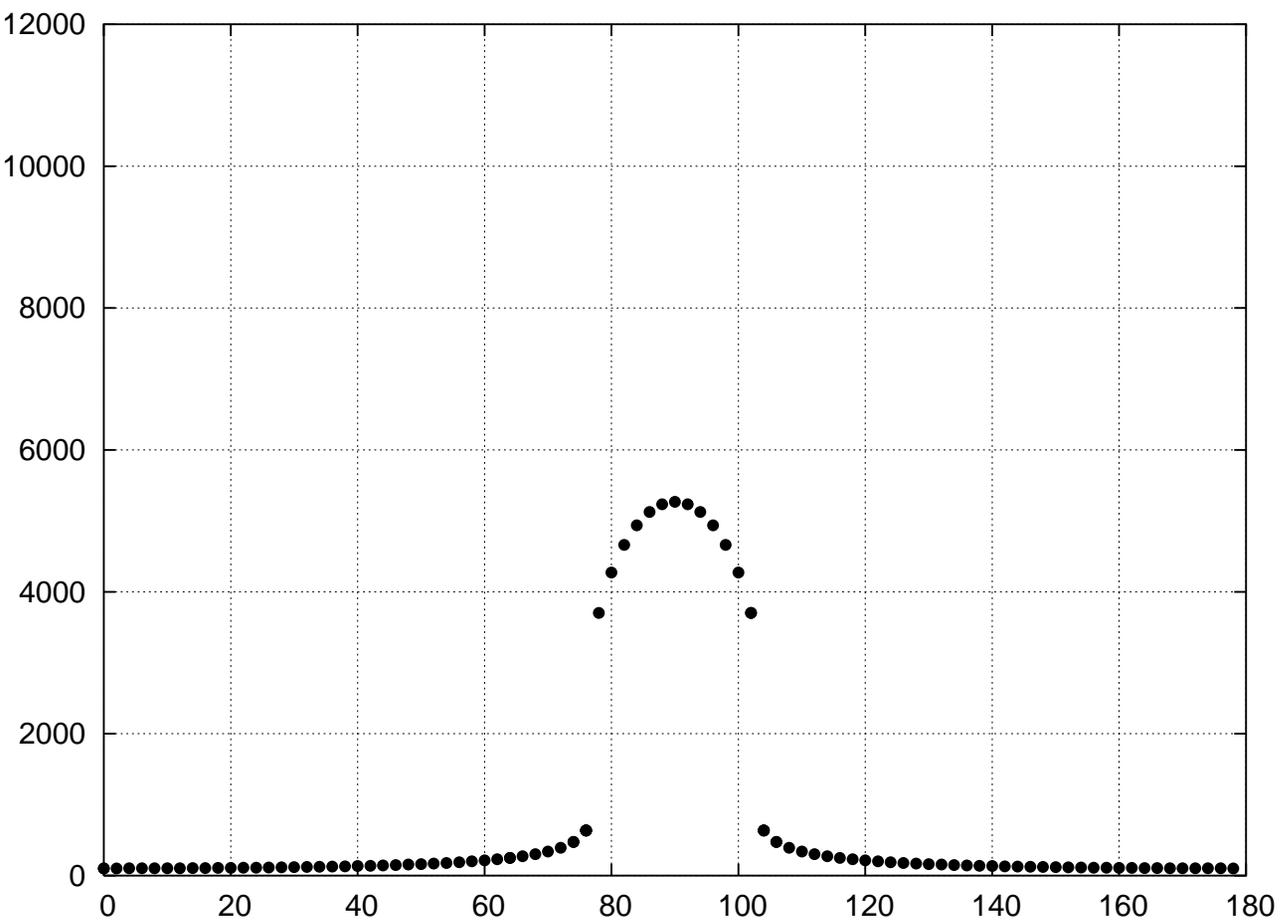}}
\caption{No anisotropy at all, $K_{u}=K_{4}=0$}
\end{figure}

\begin{figure}[h]
\epsfysize=18cm
\centerline{\epsfbox{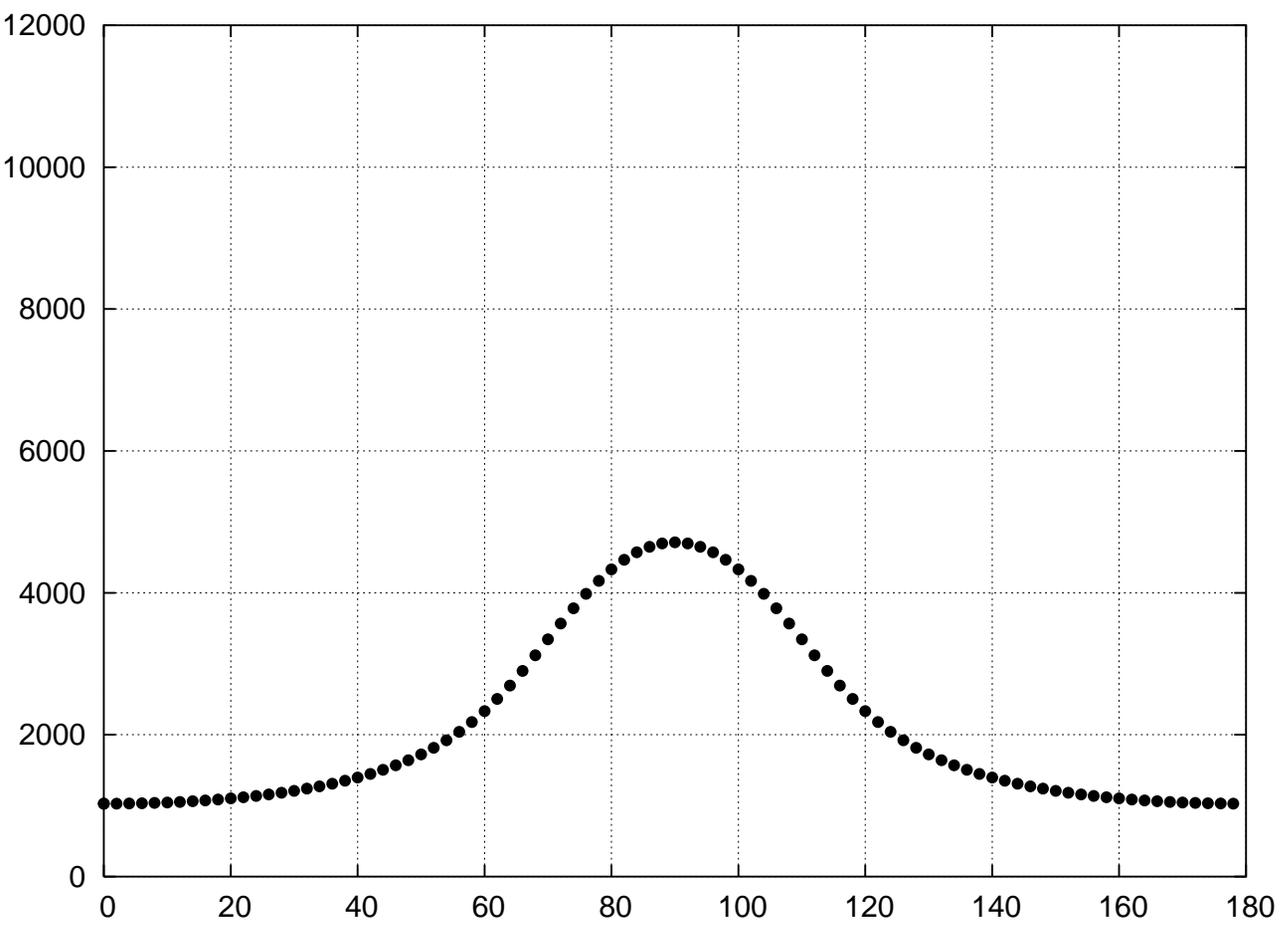}}
\caption{$K_{u}=-1.11\times 10^{5}$, $K_{4}=0$}
\end{figure}

\begin{figure}[h]
\epsfysize=18cm
\centerline{\epsfbox{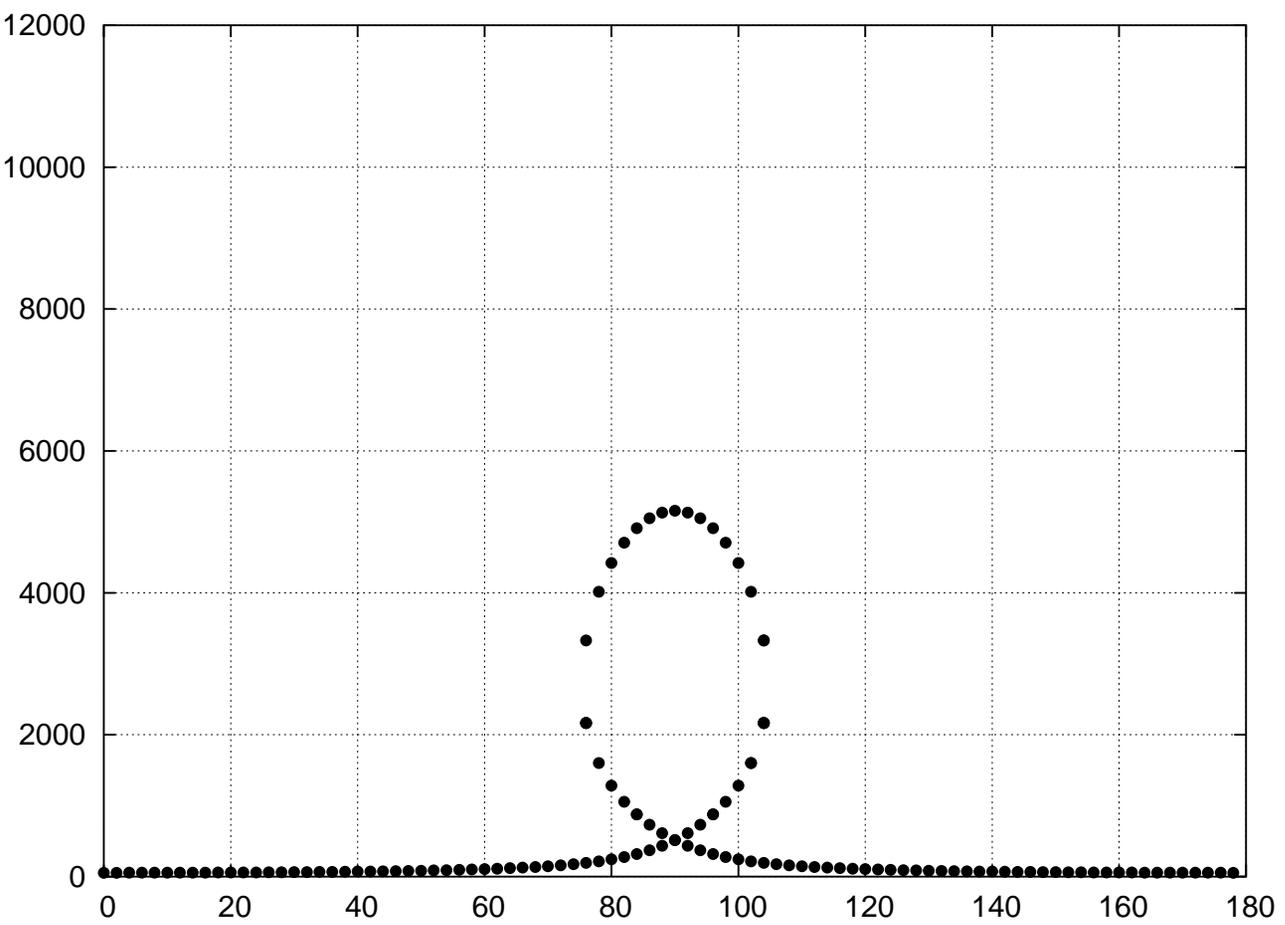}}
\caption{$K_{u}=+0.20\times 10^{5}$, $K_{4}=-K_{u}$}
\end{figure}

\begin{figure}[h]
\epsfysize=18cm
\centerline{\epsfbox{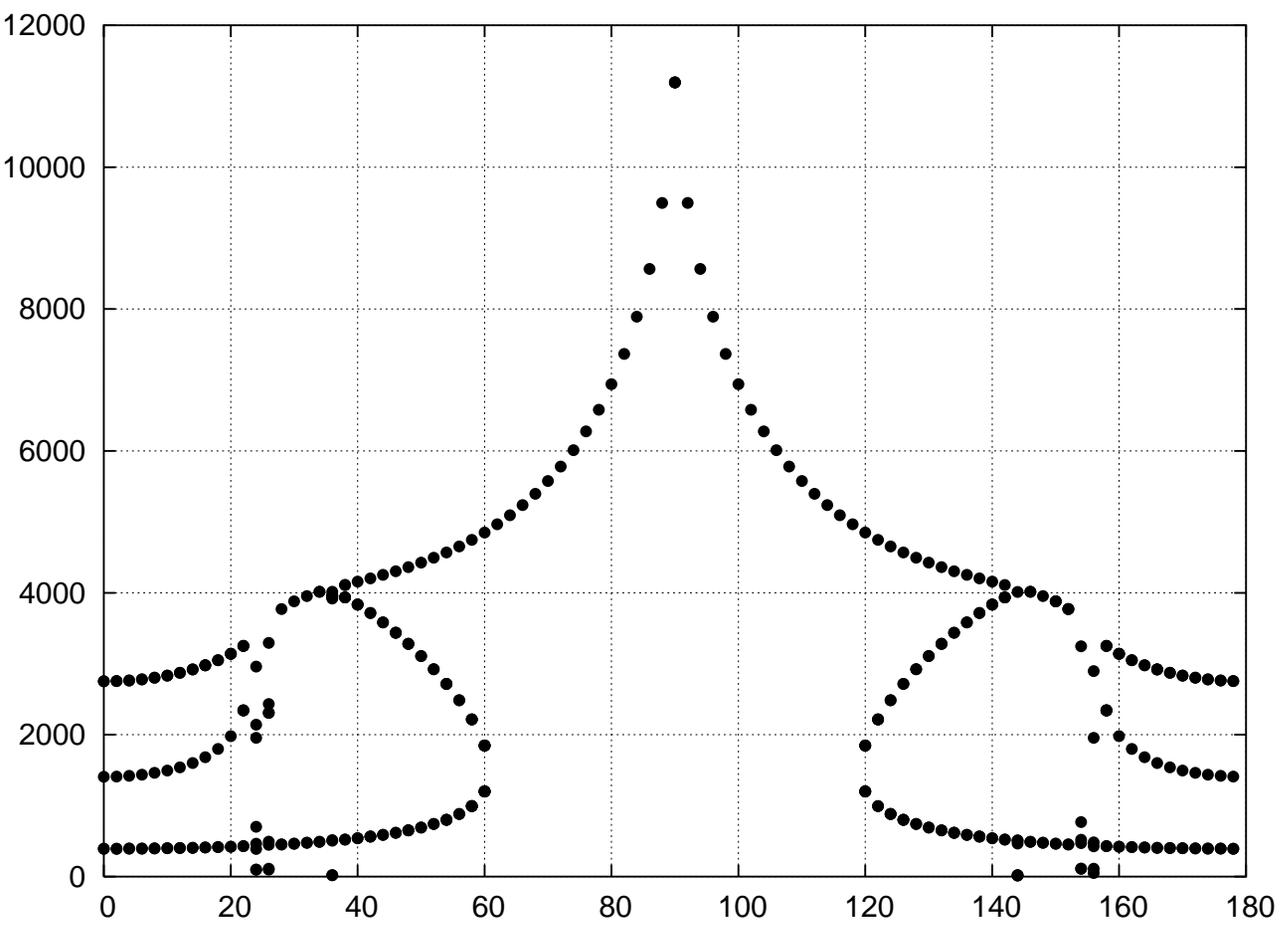}}
\caption{$K_{u}=+2.00\times 10^{5}$, $K_{4}=+4.41\times 10^{5}$}
\end{figure}

\begin{figure}[h]
\epsfysize=18cm
\centerline{\epsfbox{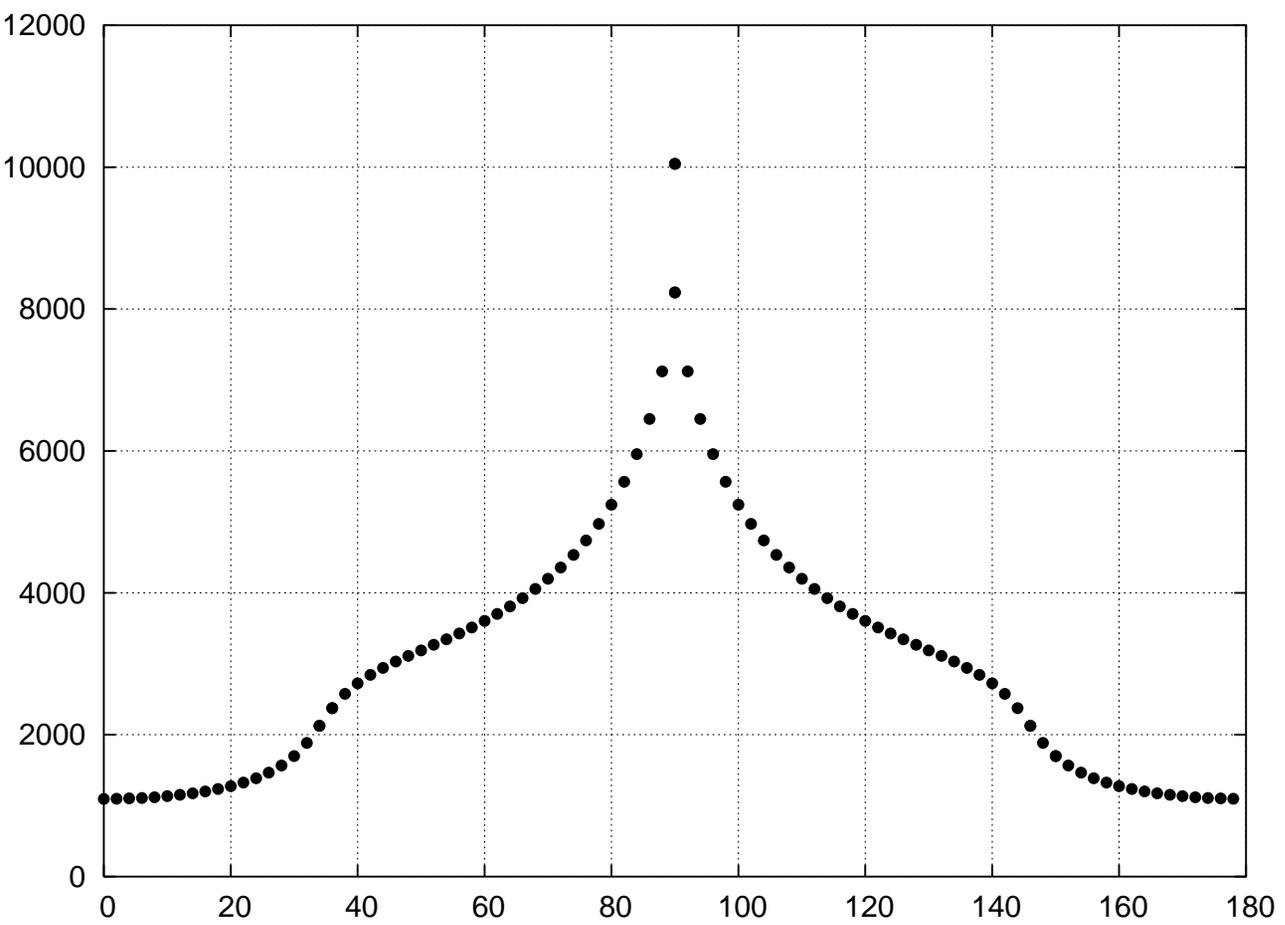}}
\caption{$K_{u}=+1.59\times 10^{5}$, $K_{4}=+2.85\times 10^{5}$}
\end{figure}

\begin{figure}[h]
\epsfysize=18cm
\centerline{\epsfbox{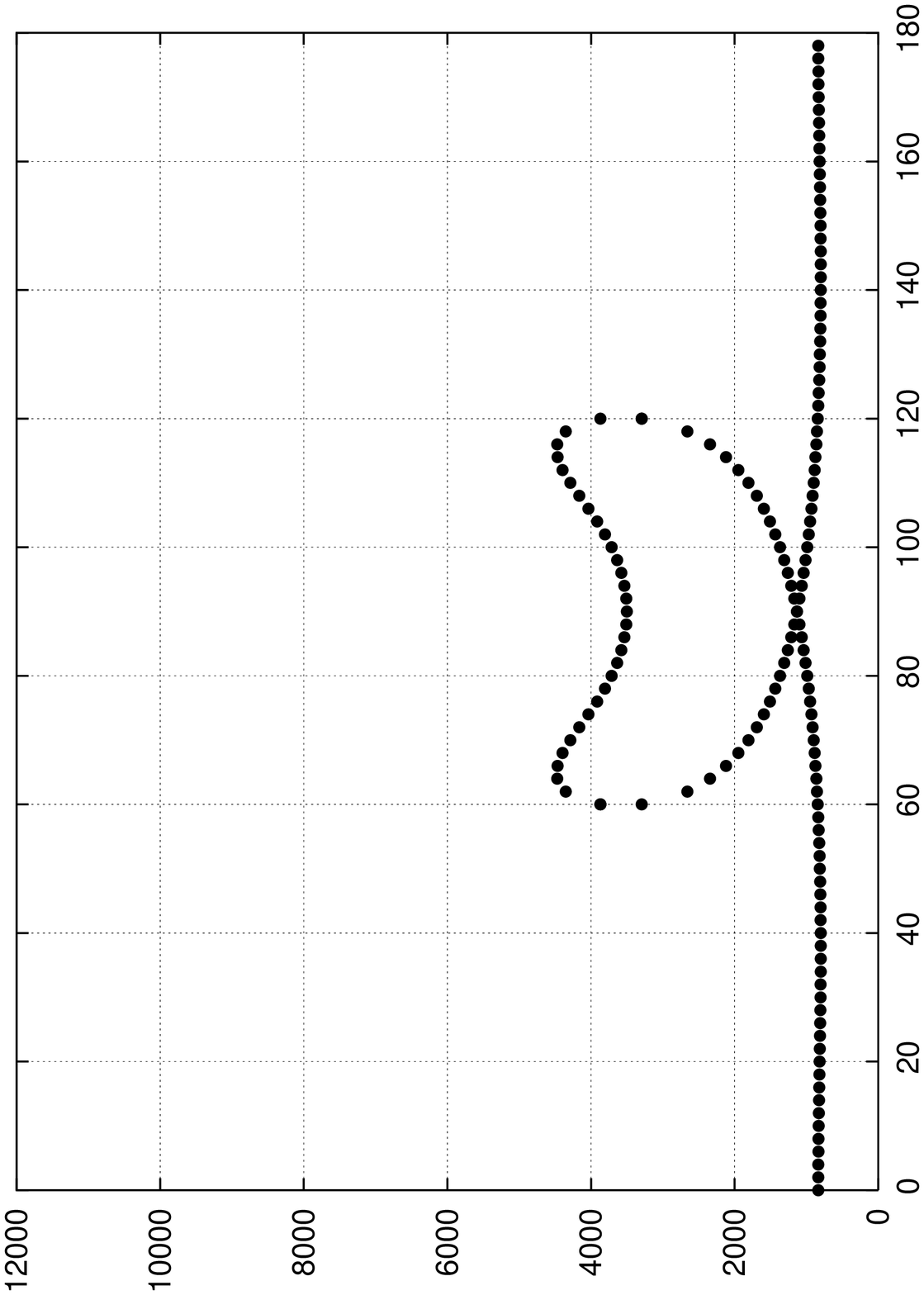}}
\caption{$K_{u}=+1.19\times 10^{5}$, $K_{4}=-2.38\times 10^{5}$}
\end{figure}

\begin{figure}[h]
\epsfysize=18cm
\centerline{\epsfbox{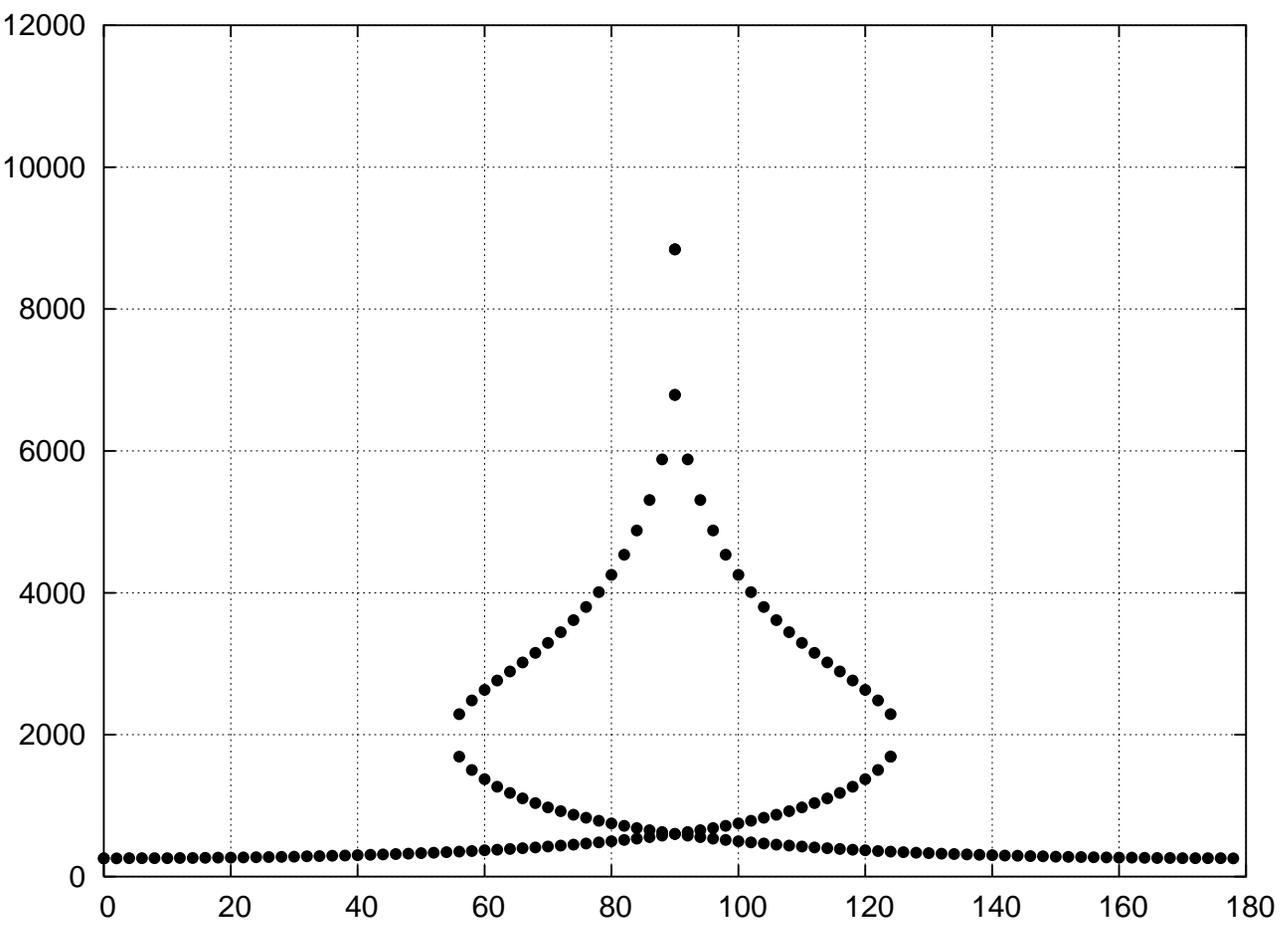}}
\caption{$K_{u}=-0.19\times 10^{5}$, $K_{4}=+2.90\times 10^{5}$}
\end{figure}

\begin{figure}[h]
\epsfysize=18cm
\centerline{\epsfbox{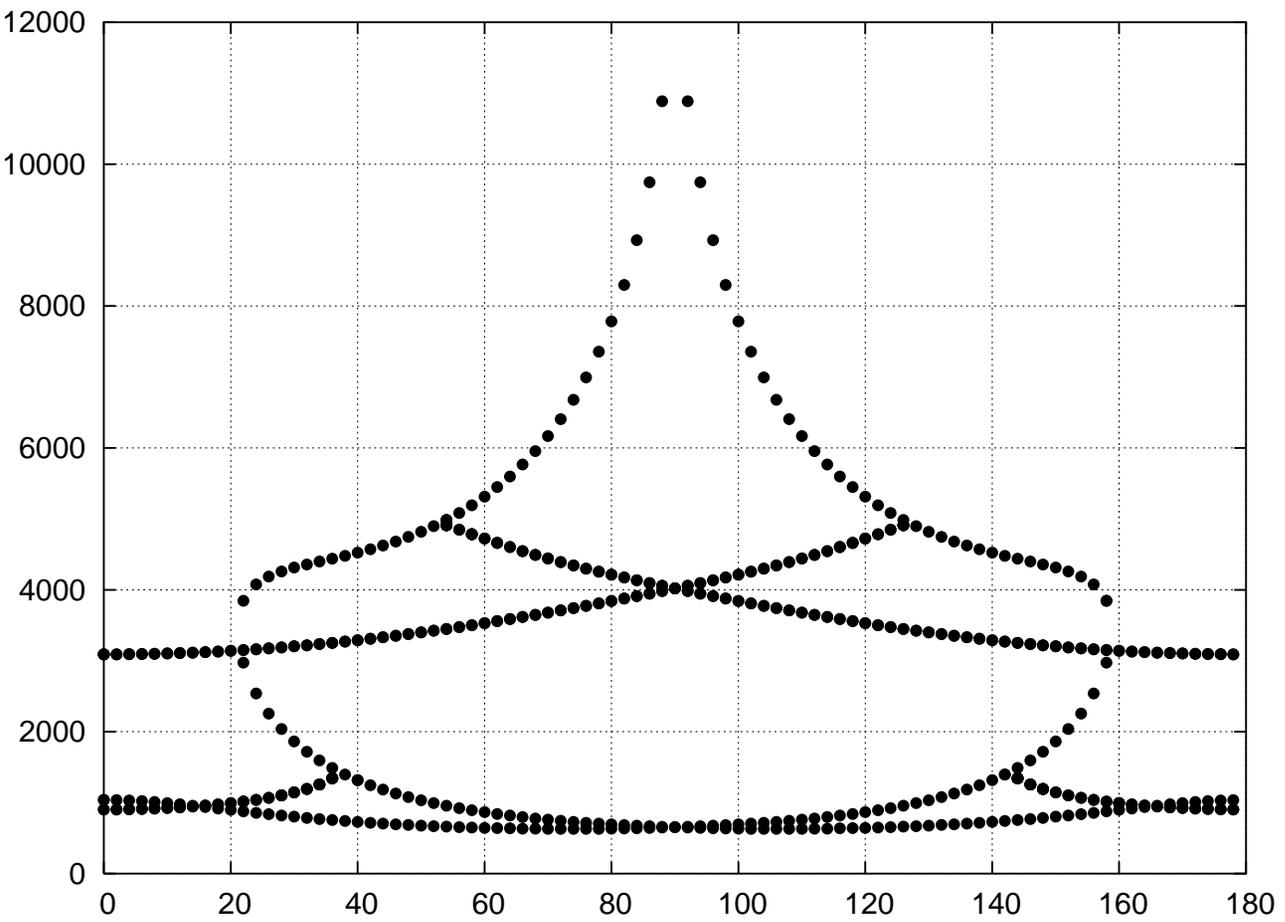}}
\caption{$K_{u}=-1.19\times 10^{5}$, $K_{4}=+7.16\times 10^{5}$}
\end{figure}

\begin{figure}[h]
\epsfysize=18cm
\centerline{\epsfbox{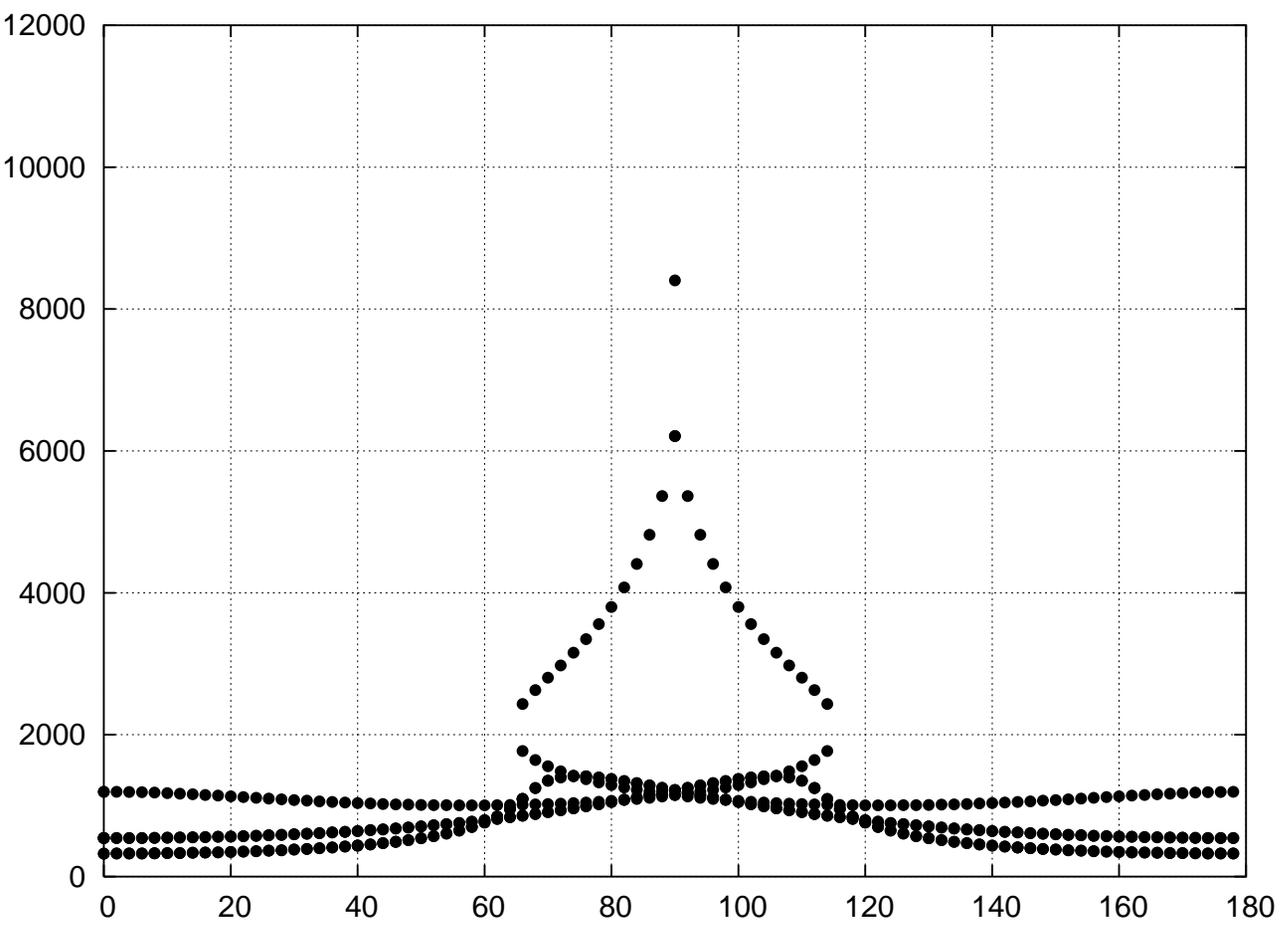}}
\caption{$K_{u}=-1.39\times 10^{5}$, $K_{4}=+3.18\times 10^{5}$}
\end{figure}

\begin{figure}[h]
\epsfysize=18cm
\centerline{\epsfbox{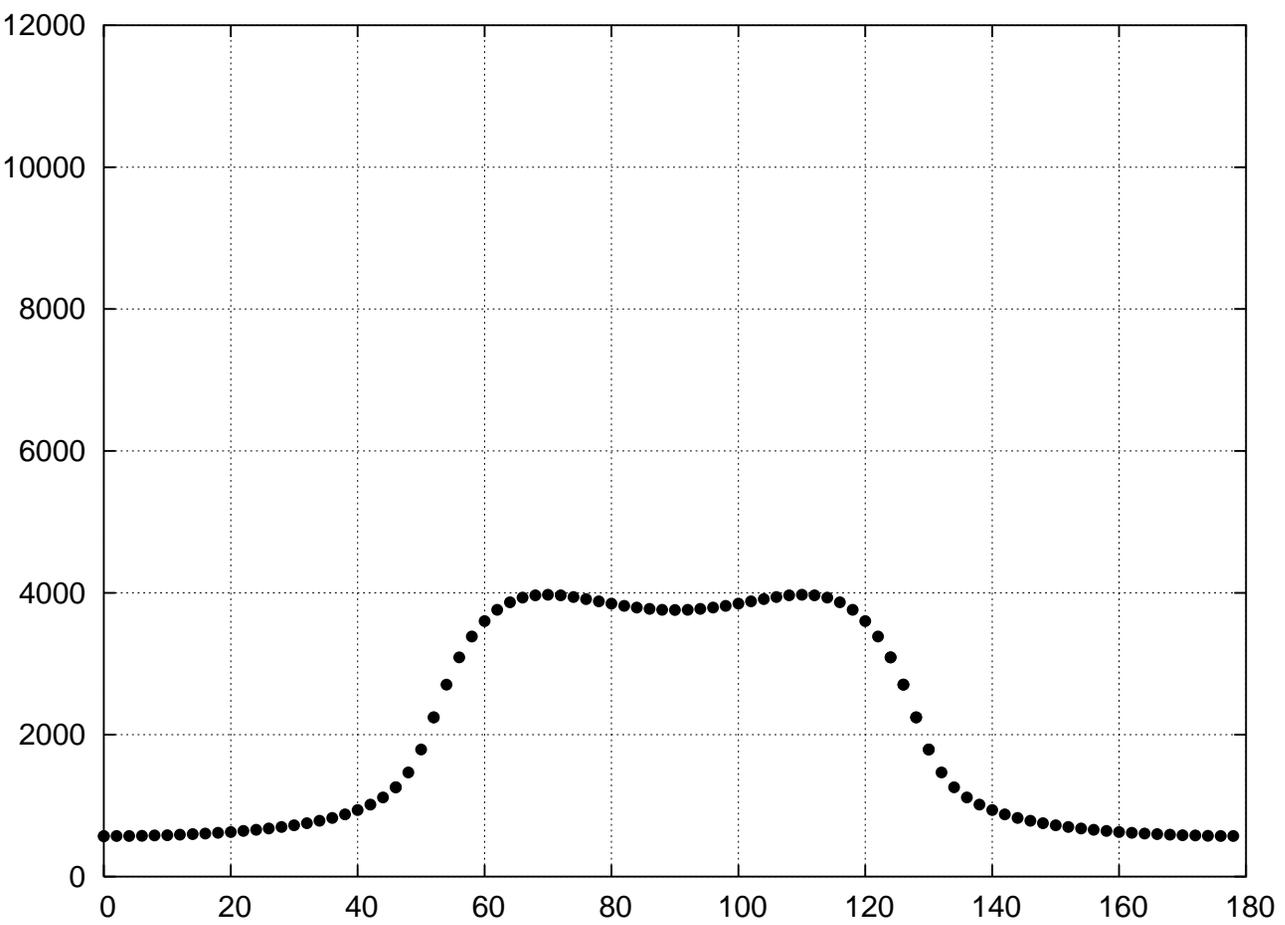}}
\caption{$K_{u}=-5.96\times 10^{5}$, $K_{4}=-1.19\times 10^{5}$}
\end{figure}

\begin{figure}[h]
\epsfysize=18cm
\centerline{\epsfbox{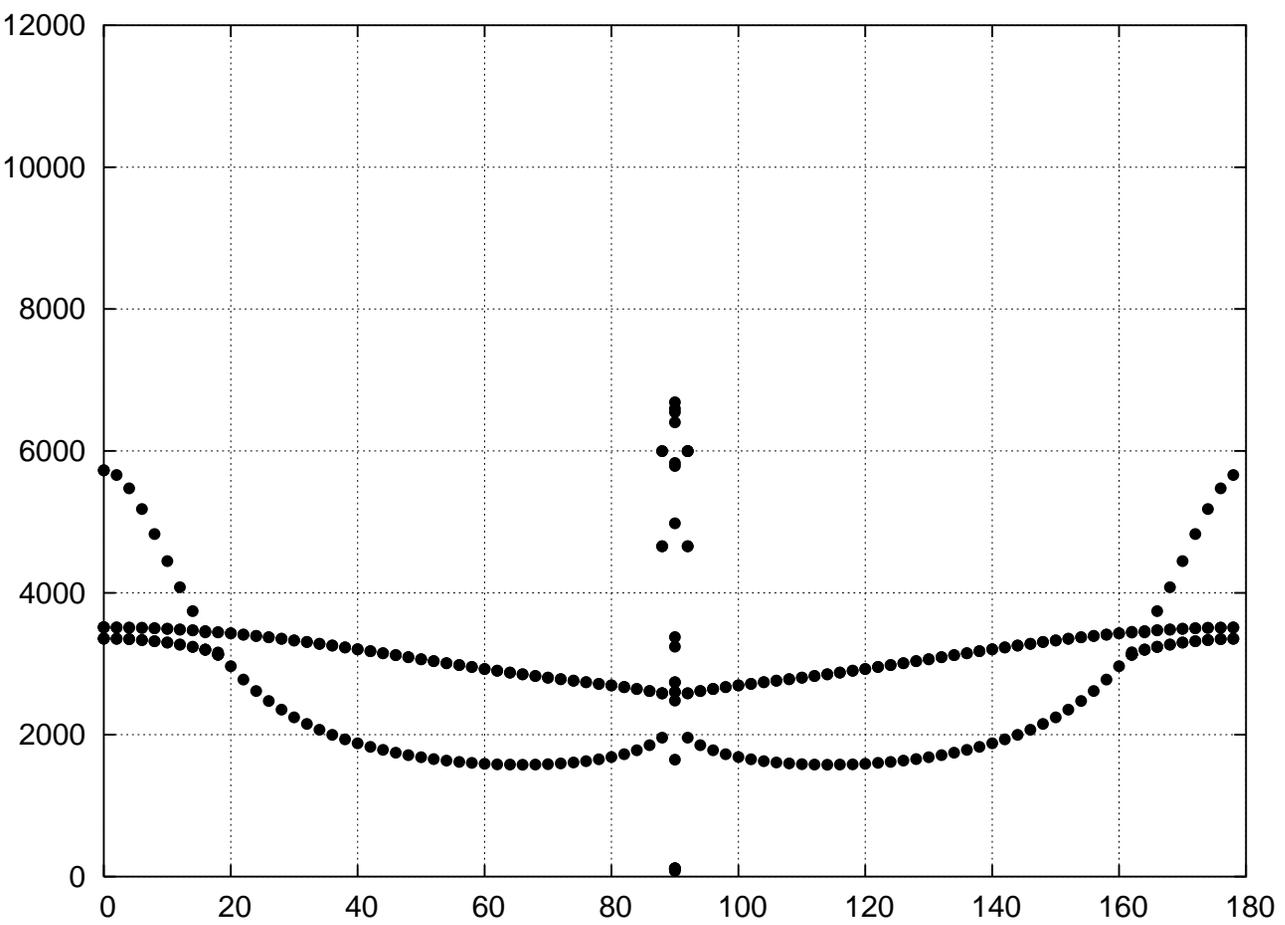}}
\caption{$K_{u}=-7.16\times 10^{5}$, $K_{4}=+4.77\times 10^{5}$}
\end{figure}

\begin{figure}[h]
\epsfysize=18cm
\centerline{\epsfbox{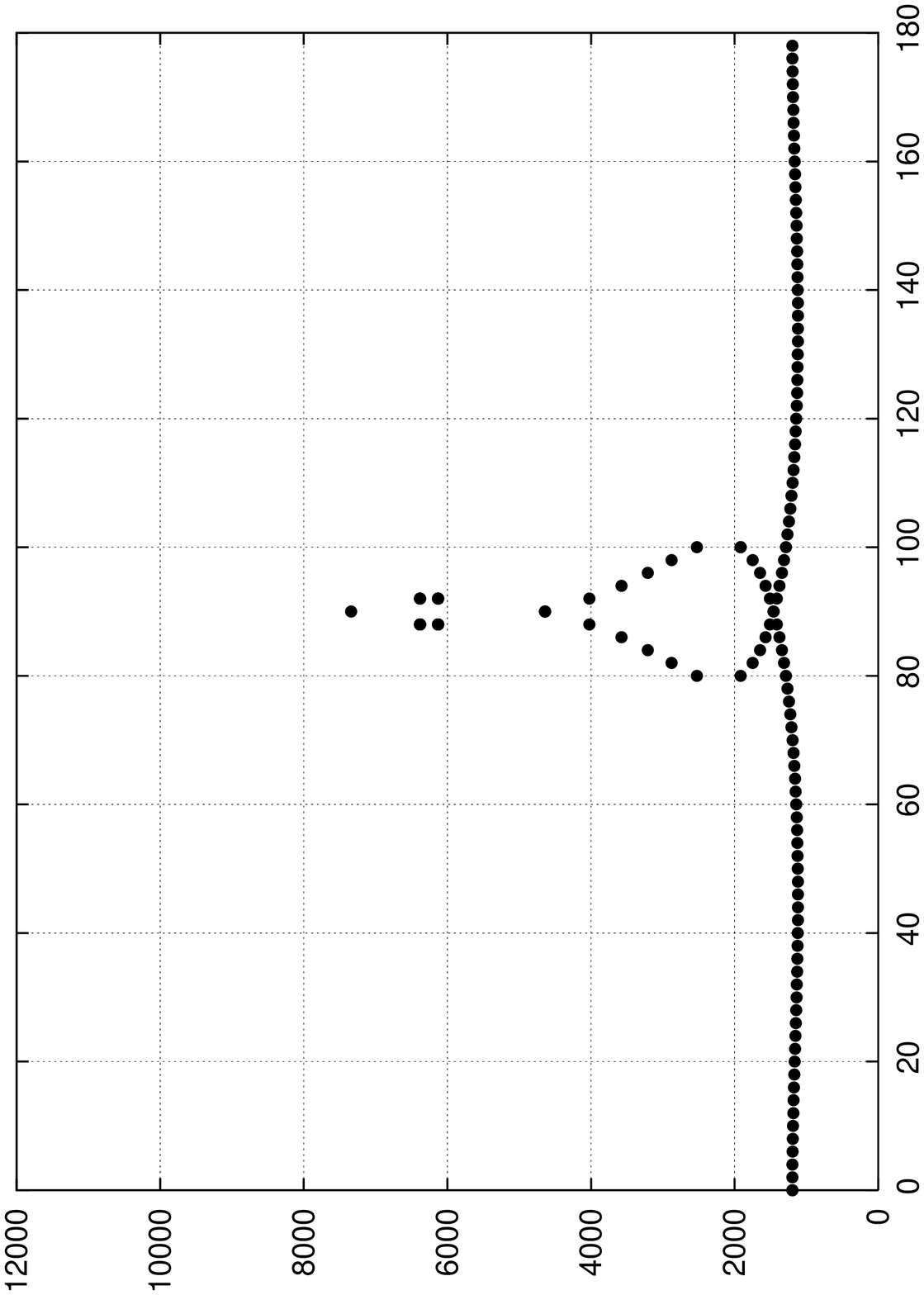}}
\caption{$K_{u}=-1.39\times 10^{5}$, $K_{4}=+2.38\times 10^{5}$}
\end{figure}

\end{document}